\documentclass{article}
\usepackage{Vmargin}
\setpapersize{A4}
\usepackage{cite}
\usepackage{amsmath}
\usepackage{amssymb}
\usepackage{epsfig}
\usepackage{subfigure}
\newcommand{\HSG}{\mathcal{H}_\mathrm{base}}
\newcommand{\Hf}{\mathcal{H}_\mathrm{field}}
\newcommand{\Ha}{\mathcal{H}}
\newcommand{\ZSG}{Z_\mathrm{base}}
\newcommand{\fSG}{f_\mathrm{base}}
\newcommand{\ff}{f_\mathrm{field}}
\newcommand{\ffRS}{f^\mathrm{(RS)}_\mathrm{field}}
\newcommand{\ffRSB}{f^\mathrm{(1RSB)}_\mathrm{field}}
\newcommand{\sech}{\operatorname{sech}}
\newcommand{\eqind}{\hspace{15mm}}
\newcommand{\fev}{f_\mathrm{E}}
\newcommand{\fod}{f_\mathrm{O}}
\renewcommand{\appendix}
   {\renewcommand{\thesection}{Appendix~\Alph{section}}
    \setcounter{section}{0}}
\title{Multispin Ising spin glasses with ferromagnetic interactions}
\author{Peter Gillin$^1$, Hidetoshi Nishimori$^2$, and David Sherrington$^1$\\
\footnotesize{$^1$ Physics Department, University of Oxford,
Theoretical Physics, 1 Keble Road, Oxford OX1 3NP, UK}\\
\footnotesize{$^2$ Department of Physics, Tokyo Institute of Technology,
Oh-Okayama, Meguro-ku, Tokyo 152-8551, Japan}}
\date{}
\begin{document}
\bibliographystyle{/home/wytham/gillin/latex/pg}
\maketitle
\begin{abstract}
\noindent
We consider the thermodynamics of an infinite-range Ising $p$-spin
glass model with an additional $r$-spin ferromagnetic interaction. For
$r=2$ there is a continuous transition to a ferromagnetic phase, while
for $r>2$ the transition is first order. We find both glassy and
non-glassy ferromagnetic phases, with replica symmetry breaking of
both the one step and full varieties. We obtain new results for the
case where $r=p>2$, demonstrating the existence of a non-glassy
ferromagnetic phase, of significance to error-correcting codes.

\end{abstract}
\section{Introduction}
Competition between quenched randomness, frustration, and a tendency
to order uniformly causes non-trivial behaviour in many-body
systems. Typical examples include spin glasses with ferromagnetic
bias\cite{mpv:book,s:rev,mydosh:expt}, neural
networks\cite{hkp:nnets}, proteins\cite{wot:protfold}, and
error-correcting codes\cite{s:ecc}. When the quenched randomness
dominates, these systems exhibit randomly frozen (spin glass)
states. When the ordering tendency dominates, they exhibit uniformly
ordered (non-glassy ferromagnetic) states. In the intermediate regime
they can exhibit mixed (glassy ferromagnetic) states. The phase
diagrams around this crossover often have a rich structure.

One of the principal purposes of this paper is to present a new method
to construct the phase diagrams of such systems from an understanding
of the related system with only the random term and an external field
term. We apply this method to an Ising spin glass with a $p$-spin
Gaussian random interaction and an $r$-spin ordering interaction, both
of infinite range. It has been shown\cite{g:ipg2sg} using replica
theory that for $p>2$, the system with neither ordering nor field
terms has two phase transitions: a discontinuous one step replica
symmetry breaking, followed by a full replica symmetry breaking at
lower temperature. Subsequent work investigated the one step
transition with the addition of either a field\cite{of:ipsgh} or an
ordering interaction with $r=p$\cite{nw:ecc}. We extend the results to
the case of general $r$ and also examine the transition from one step
to full replica symmetry breaking. This work follows on from a
previous paper\cite{gs:prsg} in which we dealt not with an Ising spin
system, but with a technically simpler spherical spin system.

The order of the paper is as follows. In \S\ref{sect:gen} we present
the general method. In \S\ref{sect:rep} we review the replica theory
of the Ising $p$-spin glass. In \S\ref{sect:res} we discuss the phase
diagrams of the Ising $p$-spin glass with $r$-spin ferromagnetism. In
\S\ref{sect:nish} we present some exact results using a gauge
transformation for the case where $r=p$. Finally, we make concluding
remarks and present technical appendices.

\section{The constrained magnetization approach to mean field
ferromagnetism \label{sect:gen}}

We consider a general system of scalar spins with a Hamiltonian
\begin{gather}
\Ha_r(J_0) = \HSG - \frac{J_0 (r-1)!}{N^{r-1}} \sum_{ i_1 < i_2 \dots
< i_r } \sigma_{i_1} \dots \sigma_{i_r}
\label{genH}
\end{gather}
where the first term is the Hamiltonian of any system (referred to as
the base system), and the second term is an infinite-range $r$-spin
ferromagnetic interaction. To leading order in $1/N$ the partition
function is
\begin{align}
Z_r(J_0) &= \sum_{\{ \sigma_i \}} \exp \left[ -\beta \HSG + N \beta J_0
\frac{1}{r} \left( \frac{1}{N} \sum_i \sigma_i \right)^r \, \right]
\\
&= \int dM\, \exp -N\beta \left( \fSG(M) - \frac{1}{r} J_0 M^r
\right)
\label{genZ}
\intertext{where}
\fSG(M) &= -\frac{1}{N\beta} \ln \ZSG(M)\,,
\\
\ZSG(M) &= \sum_{\{ \sigma_i \}} \delta \left( M - \frac{1}{N} \sum_i
\sigma_i \right) \exp -\beta \HSG\,.
\end{align}
We recognize $\fSG(M)$ as the free energy per site of the base system
when the magnetization per spin is constrained (via the
$\delta$-function in the partition function) to be $M$. Given a
knowledge of this function, we can carry out the integral over $M$ in
\eqref{genZ} to leading order in $1/N$ by the saddle point method,
giving a free energy per spin of
\begin{gather}
f_r(J_0) = -\frac{1}{N\beta} \ln Z_r(J_0) = \fSG(M_r) - \frac{1}{r}
J_0 M_r^r
\label{genfr}
\end{gather}
where the equilibrium magnetization $M_r(J_0)$ minimizes the right
hand side, and so is a solution to the equation
\begin{gather}
\fSG'(M_r) = J_0 M_r^{r-1}.
\label{gensc}
\end{gather}
We note further that if we take $r=1$, the second term of \eqref{genH}
represents an external magnetic field $h=J_0$, and so the free energy
of the constrained base system is related simply to that of the
unconstrained base system in an external magnetic field by
\begin{gather}
f_\mathrm{field}(h) = \fSG(M_\mathrm{field}(h)) - h
M_\mathrm{field}(h),
\label{genf1}
\\
f'_\mathrm{base}(M_\mathrm{field}(h)) = h\,.
\end{gather}
Therefore a knowledge of the free energy of
either system is enough to determine that of the ferromagnetic system
via \eqref{genfr} and \eqref{gensc}: the ferromagnetic term acts as an
effective field
\begin{gather}
h = J_0 M^{r-1}
\label{heff}
\end{gather}
which must be determined self-consistently.

\section{The replica theory of the Ising $p$-spin glass\label{sect:rep}}

In the remainder of this paper we shall concentrate on the example of
an infinite-ranged Ising $p$-spin glass model. The Hamiltonian we use
is
\begin{gather}
\Ha_r = \sum_{i_1 < i_2 \dots < i_p} J_{i_1 \dots i_p} \sigma_{i_1}
\dots \sigma_{i_p} - \frac{J_0 (r-1)!}{N^{r-1}} \sum_{i_1 < i_2 \dots <
i_r} \sigma_{i_1} \dots \sigma_{i_r}
\label{Hfull}
\end{gather}
where $J_{i_1 \dots i_p}$ are independent quenched random couplings
given by a Gaussian distribution with zero mean and variance $p! J^2 /
2 N^{p-1}$, and $\sigma_i = \pm 1$ are Ising spins. For $p=2$ we
recover a generalized Sherrington--Kirkpatrick
model\cite{sk:modelB,p:ansatz3} which has only full replica symmetry
breaking. In the present paper we consider $p>2$ so that there also is
a region of one step replica symmetry breaking.  In the limit $p \to
\infty$, we recover the random energy model (REM)\cite{d:rem,dw:remfm}
which has no full replica symmetry breaking: this will be discussed
further in \S\ref{sect:res}.

In the spirit of the previous section, we consider first the simpler
\begin{gather}
\Hf = \sum_{i_1 < i_2 \dots < i_p} J_{i_1 \dots i_p} \sigma_{i_1}
\dots \sigma_{i_p} - h \sum_i \sigma_i\,.
\label{Hfield}
\end{gather}
Previous studies\cite{g:ipg2sg,of:ipsgh,nw:ecc} used the replica
method\cite{mpv:book,s:rev} to obtain the free energy per site $\ff$
given by
\begin{multline}
\beta \ff = \lim_{n \to 0} \frac{1}{n} \Biggl[ \frac{1}{4} (p-1)
\beta^2 J^2 \sum_{a \neq b} q_{ab}^p -
\ln \sum_{\{ \sigma^a \}} \exp \Biggl( \frac{1}{4} p \beta^2 J^2
\sum_{a \neq b} q_{ab}^{p-1} \sigma^a \sigma^b + \beta h \sum_a
\sigma^a \Biggr) \Biggr] - \frac{\beta^2 J^2}{4}
\end{multline}
which must be extremized for the replica overlaps $q_{ab} =
\tfrac{1}{N} \sum_i \langle \sigma_i^a \sigma_i^b \rangle$ for $1
\leqslant a \neq b \leqslant n$. In the replica symmetric (RS) ansatz
($q_{ab}=q$ for all $a \neq b$) this is
\begin{multline}
\beta \ffRS = - \frac{1}{4} (p-1) \beta^2 J^2 q^p - \frac{1}{4}
\beta^2 J^2+ \frac{1}{4} p \beta^2 J^2 q^{p-1} \\ - \int
\frac{dz}{\sqrt{2\pi}}\, e^{-z^2/2} \ln 2 \cosh \left(
\sqrt{\tfrac{1}{2} p \beta^2 J^2 q^{p-1}}\, z + \beta h \right)
\label{fRS}
\end{multline}
where the overlap $q$ is given by the self-consistency equation
\begin{gather}
q = \int \frac{dz}{\sqrt{2\pi}}\, e^{-z^2/2} \tanh^2 \left(
\sqrt{\tfrac{1}{2} p \beta^2 J^2 q^{p-1}}\, z + \beta h \right).
\label{scrs}
\end{gather}
This RS solution is stable against small replica symmetry breaking
Almeida--Thouless fluctu\-ations\cite{at:skus} if
\begin{gather}
\frac{1}{2} p(p-1) \beta^2 J^2 q^{p-2} \int \frac{dz}{\sqrt{2\pi}}\,
e^{-z^2/2} \sech^4 \left( \sqrt{\tfrac{1}{2} p \beta^2 J^2 q^{p-1}}\, z
+ \beta h \right) < 1\,.
\label{atrs}
\end{gather}
The magnetization of the RS solution is
\begin{gather}
M = \int \frac{dz}{\sqrt{2\pi}}\, e^{-z^2/2} \tanh \left(
\sqrt{\tfrac{1}{2} p \beta^2 J^2 q^{p-1}}\, z + \beta h \right).
\label{Mrs}
\end{gather}

In the one step replica symmetry breaking (1RSB)
ansatz\cite{p:ansatz2} the free energy per site is $\ffRSB$ given by
\begin{multline}
\beta \ffRSB = - \frac{1}{4} (p-1) \beta^2 J^2 \left[ (1-x) q_1^p + x
q_0^p \right] - \frac{1}{4} \beta^2 J^2 + \frac{1}{4} p \beta^2 J^2
q_1^{p-1} \\ - \frac{1}{x} \int \frac{dz_0}{\sqrt{2\pi}}\,
e^{-z_0^2/2} \ln 2^x \int \frac{dz_1}{\sqrt{2\pi}}\, e^{-z_1^2/2}
\cosh^x G
\label{fRSB}
\end{multline}
where
\begin{gather}
G \doteq \sqrt{\tfrac{1}{2} p \beta^2 J^2 q_0^{p-1}} \, z_0 +
\sqrt{\tfrac{1}{2} p \beta^2 J^2 (q_1^{p-1} - q_0^{p-1})} \, z_1 +
\beta h\,,
\end{gather}
and the mutual overlap $q_0$, the self overlap $q_1$, and the break
point $x$ (i.e. the weight of $q_0$ in the averaged overlap
distribution function) are given by the self-consistency equations
\begin{subequations}
\label{scrsb}
\begin{align}
q_0 &= \int \frac{dz_0}{\sqrt{2\pi}}\, e^{-z_0^2/2} \left( \frac{ \int
\frac{dz_1}{\sqrt{2\pi}}\, e^{-z_1^2/2} \cosh^x G \, \tanh
G} { \int \frac{dz_1}{\sqrt{2\pi}}\, e^{-z_1^2/2} \cosh^x
G} \right)^2,
\label{scrsb0}
\\
q_1 &= \int \frac{dz_0}{\sqrt{2\pi}}\, e^{-z_0^2/2} \; \frac{ \int
\frac{dz_1}{\sqrt{2\pi}}\, e^{-z_1^2/2} \cosh^x G \, \tanh^2
G} { \int \frac{dz_1}{\sqrt{2\pi}}\, e^{-z_1^2/2} \cosh^x
G},
\label{scrsb1}
\\
\frac{1}{4} (p-1) \beta^2 J^2 (q_1^p - q_0^p) &= - \frac{1}{x^2} \int
\frac{dz_0}{\sqrt{2\pi}}\, e^{-z_0^2/2} \, \ln \int
\frac{dz_1}{\sqrt{2\pi}}\, e^{-z_1^2/2} \cosh^x G \notag \\& \eqind
+ \frac{1}{x} \int \frac{dz_0}{\sqrt{2\pi}}\, e^{-z_0^2/2}
\; \frac{ \int \frac{dz_1}{\sqrt{2\pi}}\, e^{-z_1^2/2} \cosh^x
G \, \ln \cosh G} { \int \frac{dz_1}{\sqrt{2\pi}}\,
e^{-z_1^2/2} \cosh^x G}.
\label{scrsbx}
\end{align}
\end{subequations}
This 1RSB solution is stable against further replica symmetry breaking
fluctuations if
\begin{gather}
\frac{1}{2} p(p-1) \beta^2 J^2 q_1^{p-2} \int
\frac{dz_0}{\sqrt{2\pi}}\, e^{-z_0^2/2} \; \frac{ \int
\frac{dz_1}{\sqrt{2\pi}}\, e^{-z_1^2/2} \cosh^{x-4} G } { \int
\frac{dz_1}{\sqrt{2\pi}}\, e^{-z_1^2/2} \cosh^x G} < 1\,.
\label{atrsb}
\end{gather}
Where the 1RSB solution becomes unstable, we expect a continuous
transition to a solution given by the full Parisi ansatz, which we do
not consider in detail here. The magnetization of the 1RSB solution is
\begin{gather}
M = \int \frac{dz_0}{\sqrt{2\pi}}\, e^{-z_0^2/2} \; \frac{ \int
\frac{dz_1}{\sqrt{2\pi}}\, e^{-z_1^2/2} \cosh^x G \, \tanh
G} { \int \frac{dz_1}{\sqrt{2\pi}}\, e^{-z_1^2/2} \cosh^x
G}.
\label{Mrsb}
\end{gather}

\section{Full results and phase diagrams\label{sect:res}}

\begin{figure}
\begin{center}
\subfigure[Field case (equivalent to $r=1$)]
{\epsfig{file=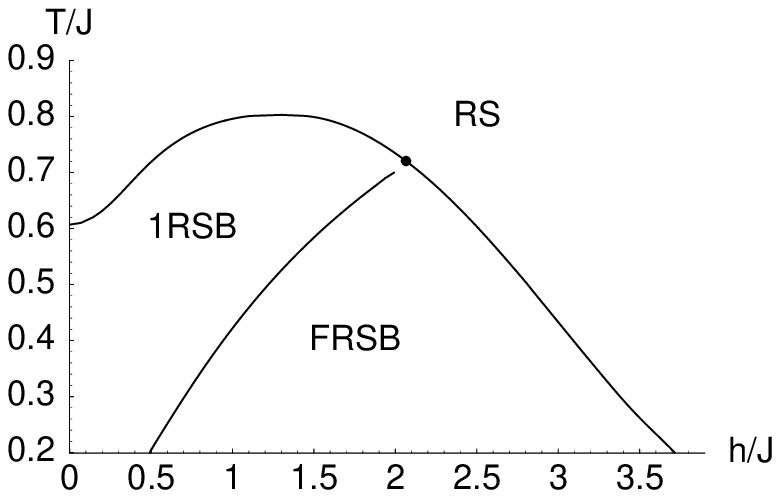,width=7cm}\label{figure:phd1}}
\quad
\subfigure[$r=2$]
{\epsfig{file=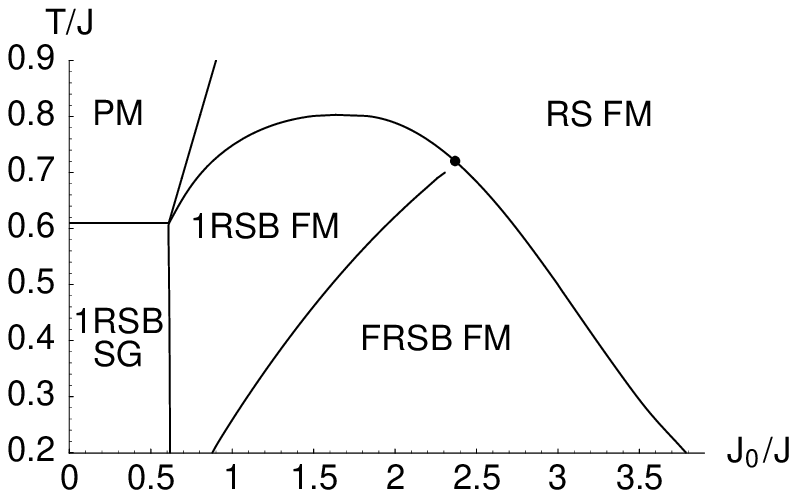,width=7cm}\label{figure:phd2}}
\\
\subfigure[$r=3$]
{\epsfig{file=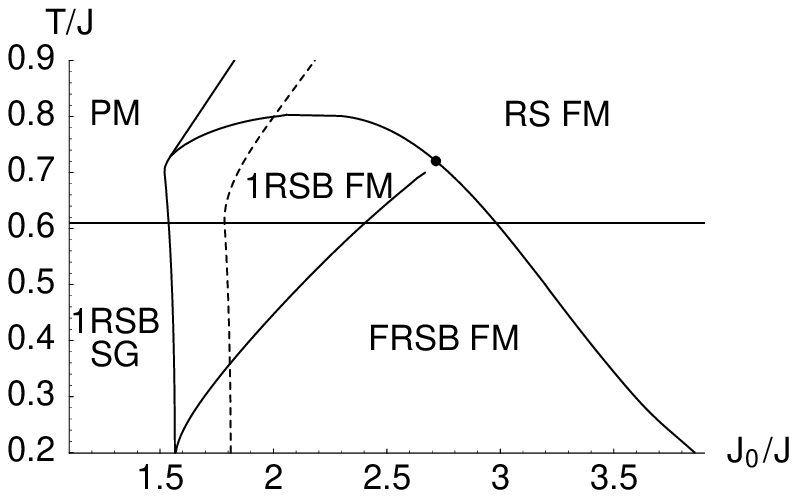,width=7cm}\label{figure:phd3}}
\quad
\subfigure[$r=4$]
{\epsfig{file=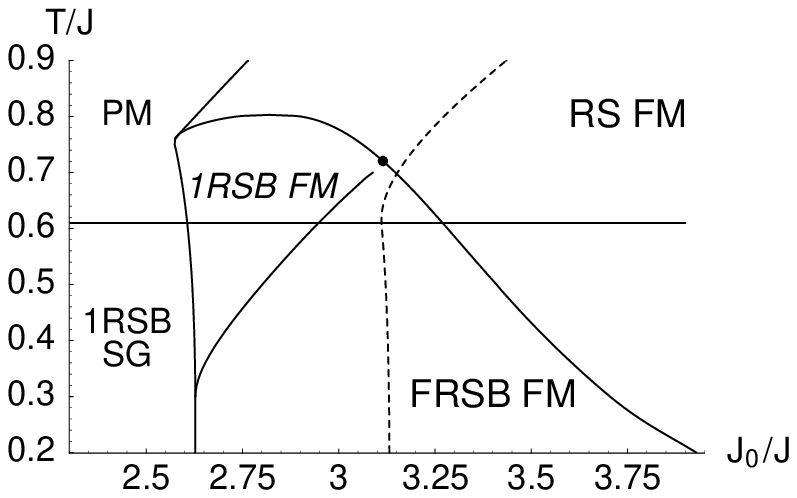,width=7cm}}
\\
\subfigure[$r=5$]
{\epsfig{file=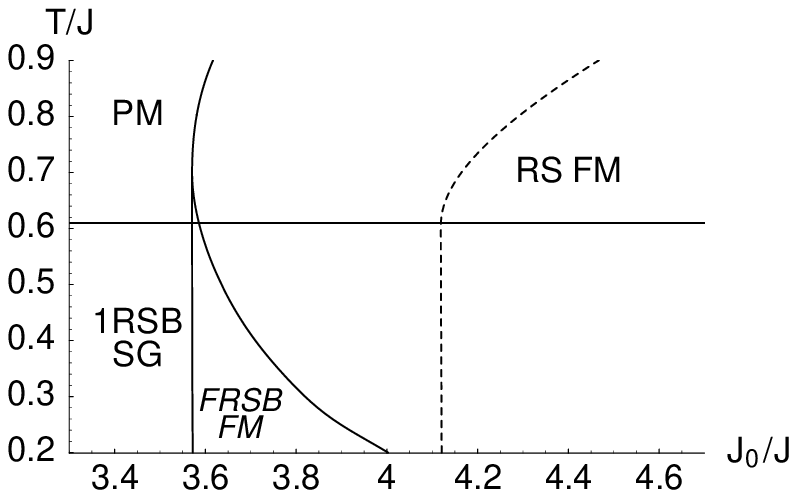,width=7cm}\label{figure:phdp}}
\quad
\subfigure[$r=6$]
{\epsfig{file=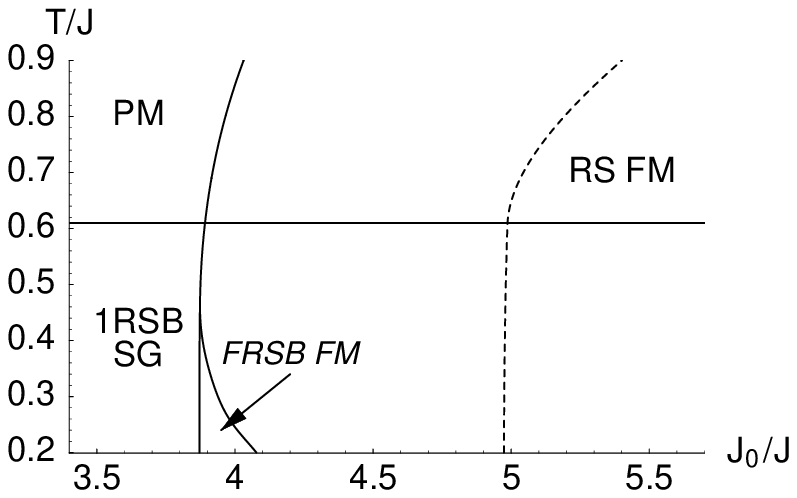,width=7cm}\label{figure:phdlast}}
\caption{The phase diagrams calculated in \S\ref{sect:res} for the
systems given by (a) the Hamiltonian \eqref{Hfield} with $p=5$, and
(b)--(f) the Hamiltonian \eqref{Hfull} with $p=5$ and
$r=2,\dots,6$. The phases are labelled by their symmetry breaking (RS,
1RSB, and FRSB) and, for $r \geqslant 2$, by their ferromagnetism
($M=0$ for the paramagnetic (PM) and spin glass (SG) phases, while $M
\neq 0$ for the ferromagnetic (FM) phases). For $r>2$, where the
ferromagnetic transition is first order, the spinodal transition is
shown with a solid line, and the thermodynamic transition with a
dashed line; phases labelled in italics are metastable. The RSB triple
point (found by the perturbative calculation of \ref{app:pert}) is
shown as a circle. (The transition to FRSB is at a temperature too low
to appear for small $h$ or $J_0$, in particular within the spin glass
phase where the effective field vanishes.)}
\label{figure:phds}
\end{center}
\end{figure}

The phase diagrams for the Hamiltonian \eqref{Hfull} were found in two
stages. They are shown for $p=5$ and $r=1,\dots,6$ in
figure~\ref{figure:phds}.

Firstly, the free energy and magnetization of the system in a field
but without ferromagnetism, as given by \eqref{Hfield}, were
found. The RS solution was found by numerical solution of
\eqref{scrs}, the 1RSB by numerical solution of \eqref{scrsb}. No
attempt was made to determine the solution with full replica symmetry
breaking (FRSB) in the Parisi scheme, but the onset of unstable modes
in the 1RSB solution, as given by \eqref{atrsb}, is taken to signify a
continuous transition to FRSB.

\begin{figure}
\begin{center}
\subfigure[$T/J = 0.55$]
{\epsfig{file=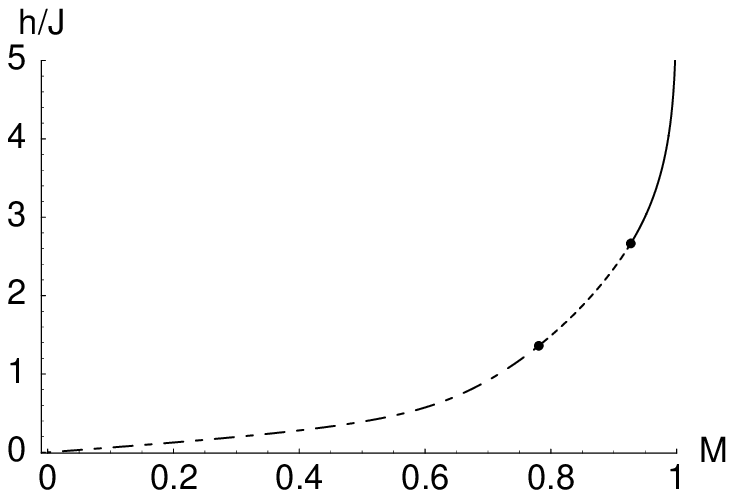,width=7cm}}
\quad
\subfigure[$T/J = 0.65$]
{\epsfig{file=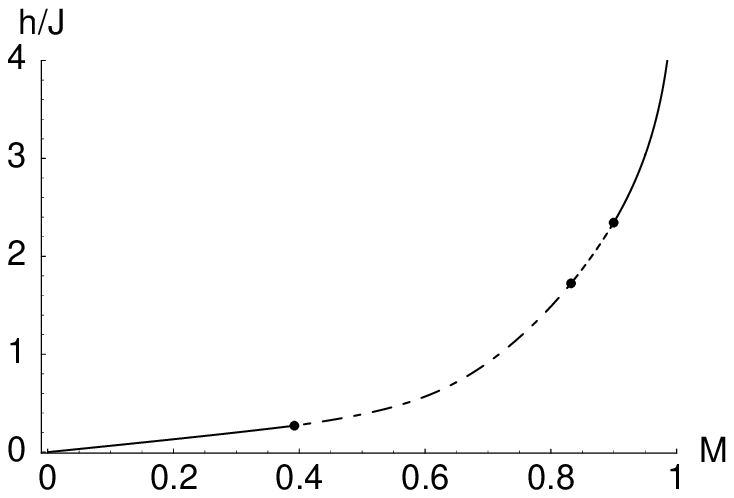,width=7cm}}
\\
\subfigure[$T/J = 0.75$]
{\epsfig{file=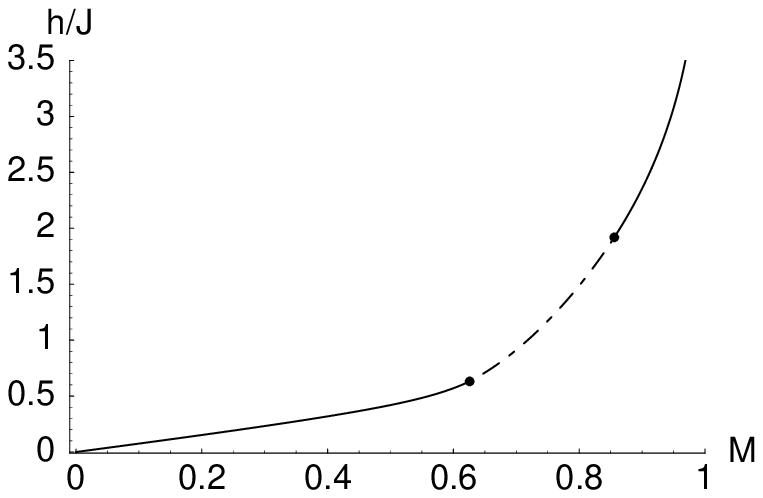,width=7cm}}
\quad
\subfigure[$T/J = 0.85$]
{\epsfig{file=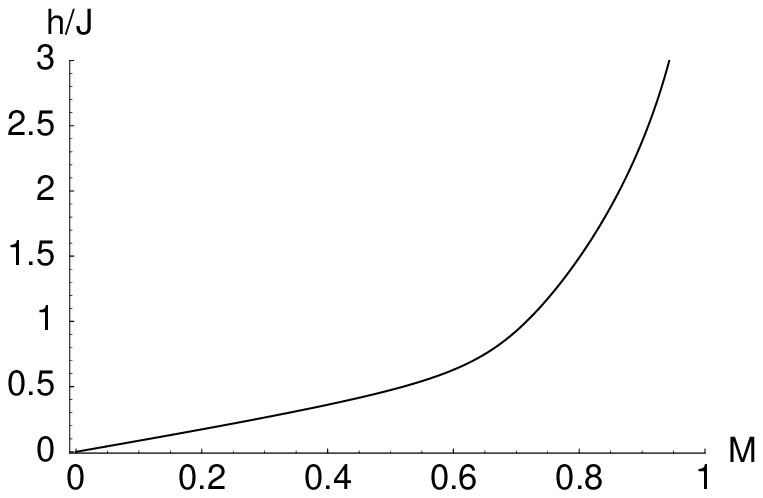,width=7cm}}
\caption{The field $h$ as a function of the magnetization $M$ for the
system given by the Hamiltonian \eqref{Hfield} with $p=5$, as
calculated in \S\ref{sect:res}, at various temperatures. The solid
curves show RS solutions, the dot-dashed curves 1RSB, and the dashed
curves FRSB; the points mark the transitions. The field here is
identical to the effective field $J_0 M^{r-1}$ in the system with
Hamiltonian \eqref{Hfull}, as discussed in \S\ref{sect:gen}.}
\label{figure:hofms}
\end{center}
\end{figure}

The resulting phase diagram for $p=5$ is plotted in
figure~\ref{figure:phd1}. This Hamiltonian is equivalent to
\eqref{Hfull} with $p=5$, $r=1$, and $J_0=h$. There are three phases:
an RS paramagnet, a 1RSB spin glass, and an FRSB spin glass. The
replica symmetry breaking transition curve has been found
previously\cite{of:ipsgh}. Where the field is less than a critical
value, there is a discontinuous one step RSB transition (D1RSB) where
$x=1$, and the transition temperature rises with $h$. Above the
critical value, the transition is continuous, first continuous one
step (C1RSB) where $q_0=q_1$, and then FRSB; and the transition
temperature falls with $h$, vanishing asymptotically as $h \to
\infty$. (Note that the D1RSB transition is continuous in the
thermodynamic sense.) The numerical calculation of the 1RSB/FRSB curve
is troublesome near the C1RSB transition line: this is because the
difference between the left and right hand sides of \eqref{atrsb} is
very small, of the order of $(q_1-q_0)^4$. Consequently we do not have
reliable results in this region, and a small gap is left in the plot
to indicate this. However, the point where this curve meets the C1RSB
line (which we shall call the RSB triple point, since RS, 1RSB, and
FRSB phases meet at it) has been determined accurately using the RS
solution and a perturbative calculation, details of which are given in
\ref{app:pert}. The magnetization $M$ was calculated with \eqref{Mrs}
and \eqref{Mrsb}, and curves of $h$ as a function of $M$ at fixed
temperature for $p=5$ are shown in figure~\ref{figure:hofms}. We
indicate where the solution is of each type (RS, 1RSB, FRSB), but note
that $h$ is not exact in the FRSB region, since it is approximated by
the 1RSB value.

Secondly, the methods of \S\ref{sect:gen} were used to calculate the
transitions in the full Hamiltonian, essentially by finding the stable
solutions of \eqref{heff} with the $h(M)$ found previously. (In the
FRSB region, the use of the 1RSB values introduces a small error in
the ferromagnetic transition lines.) The resulting phase diagrams for
$p=5$ and $r=2,\dots,6$ are plotted in
figures~\ref{figure:phd2}--\ref{figure:phdlast}.

For $r=2$ and $p>2$, the phase diagram contains six regions. For small
$J_0$, there are three $M=0$ phases: on cooling, the RS paramagnet
undergoes first a D1RSB transition, and then an FRSB (not shown
explicitly in figure~\ref{figure:phds}). Increasing $J_0$, the $M=0$
phases become unstable, and a continuous transition to $M \neq 0$
occurs. There are three ferromagnetic phases: RS, 1RSB, and
FRSB. Within the ferromagnetic region, the RS/1RSB curve has D1RSB and
C1RSB parts, as for $r=1$. (The RS/1RSB transition in the $M=0$ region
is always D1RSB, since the effective field $J_0 M^{r-1}$ vanishes.)

For $r>2$, the ferromagnetic transition is first order: on increasing
$J_0$, there is first a spinodal transition, at which a metastable $M
\neq 0$ solution appears, followed by a thermodynamic transition, at
which the comparative free energies make it the favoured state.

\begin{figure}
\begin{center}
\subfigure[$r<r_1$]
{\epsfig{file=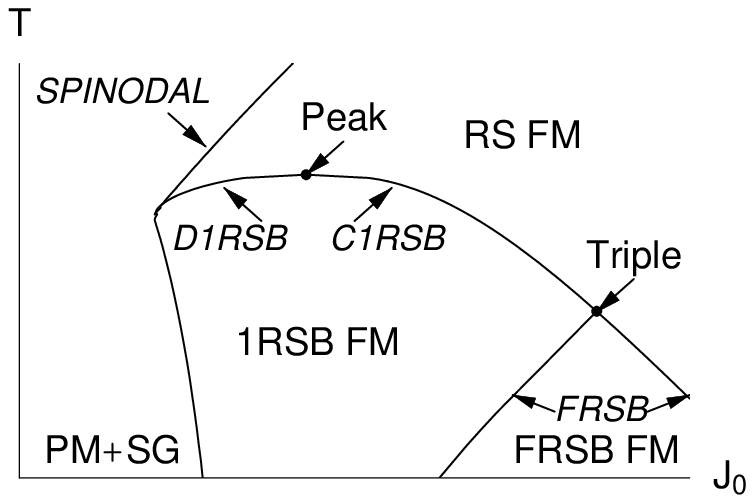,width=8cm}\label{figure:sketch1}}
\\
\subfigure[$r_1<r<r_2$]
{\epsfig{file=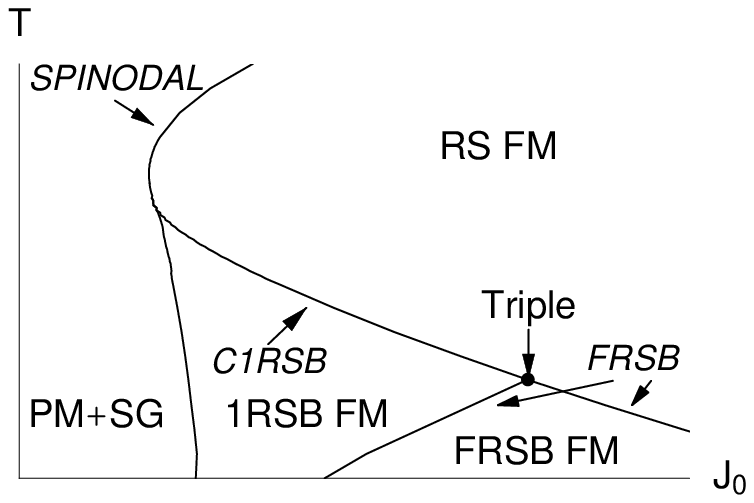,width=8cm}\label{figure:sketch2}}
\\
\subfigure[$r>r_2$]
{\epsfig{file=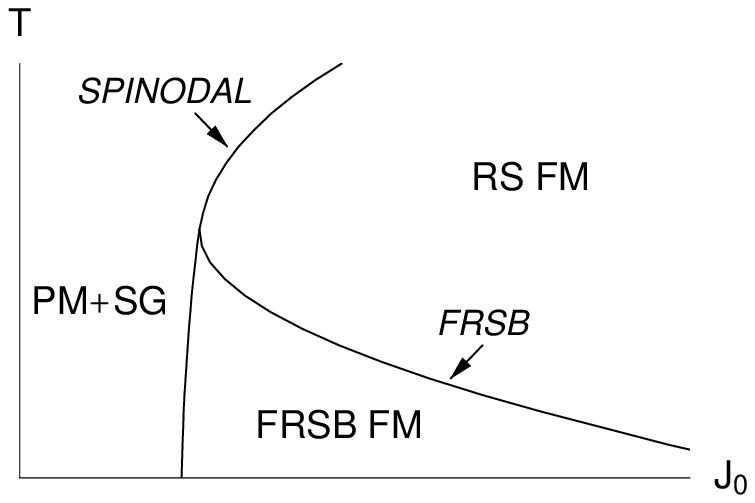,width=8cm}\label{figure:sketch3}}
\end{center}
\caption{Phase diagrams for the system given by the Hamiltonian
\eqref{Hfull}, illustrating the qualitative changes on increasing $r$
at fixed $p$. The ferromagnetic phases are labelled according to their
replica symmetry breaking. For clarity, the paramagnetic and spin
glass phases are amalgamated and the RSB transitions in the $M=0$
solution are not shown. The spinodal transition and the RSB
transitions in the ferromagnetic phase are labelled in italics. The
peak in the RS AT line and the RSB triple point are also labelled
where present. On increasing $r$, both these points move towards the
spinodal line.}
\label{figure:sketches}
\end{figure}

Increasing $r$ at fixed $p$, we identify two qualitative changes,
illustrated in figure~\ref{figure:sketches}. For $r$ slightly greater
than $2$, the diagram has the form of figure~\ref{figure:sketch1}. On
increasing $r$, the peak of the RS/1RSB curve approaches the spinodal
ferromagnetic transition line, and the D1RSB curve between them
shortens. The peak reaches the spinodal line for $r=r_1$, and there is
no D1RSB curve in the ferromagnetic region for $r \geqslant r_1$, as
in figure~\ref{figure:sketch2}. The RSB triple point approaches the
spinodal ferromagnetic transition line on increasing $r$, and the 1RSB
region shrinks. The triple point reaches the spinodal line at $r=r_2$,
and there is no ferromagnetic 1RSB region for $r \geqslant r_2$, as in
figure~\ref{figure:sketch3}. These special values of $r$ may be
evaluated for given $p$ with some accuracy, as discussed in
\ref{app:spin}. Some results are shown in table~\ref{tab:r}. They are
in accordance with an exact result, to be proved in \S\ref{sect:nish},
that $r_1$ cannot exceed $p$.

\begin{table}
\begin{center}
\begin{tabular}{|c|c|c|c|c|c|c|}
\hline
$p$ & 3 & 4 & 5 & 6 & 8 & 10
\\
\hline
$r_1$ & 2.92 & 3.82 & 4.72 & 5.62 & 7.44 & 9.30
\\
\hline
$r_2$ & 3.01 & 3.98 & 4.95 & 5.91 & 7.83 & 9.76
\\
\hline
\end{tabular}
\end{center}
\caption{Special values of $r$ for various choices of $p$. There is no
D1RSB curve in the ferromagnet for $r \geqslant r_1$, and no 1RSB
region in the ferromagnet for $r \geqslant r_2$.}
\label{tab:r}
\end{table}

Of particular interest is the behaviour of the model in the case
$r=p$; this is relevant to error-correcting codes\cite{s:ecc}, and is
amenable to an exact calculation on the Nishimori line. We observe
that there is a first order transition to a ferromagnetism which is
glassy for small enough values of $J_0$ and $T$, and that there is a
triple point where the spinodal ferromagnetic transition line and the
glassy transition line within the ferromagnet meet with infinite
slope. (Exact results in the $r=p$ case, including the fact that the
solution is RS on the line $p \beta J^2 = 2J_0$, are given in
\S\ref{sect:nish}.) The RSB region shrinks as $r=p$ increases. The
existence of an RS ferromagnetic phase persists in the limit $r=p \to
\infty$. Our results therefore contradict conclusions of Dorlas and
Wedagedera\cite{dw:remfm}, who claim that the ferromagnetic region in
the model with $r=p \to \infty$ is entirely glassy. This discrepancy
is due to the order in which the limits $r \to p$ and $p \to \infty$
are taken. The approach of Dorlas and Wedagedera is to consider the
REM with general $r$, effectively taking the limit $p \to \infty$ in
the Ising model, and to subsequently take the limit $r \to \infty$. In
doing this, they obtain results which would apply to $p \gg r \to
\infty$. In our approach we set $r=p$ and then take $p \to \infty$,
obtaining results which apply to $r=p \to \infty$, the appropriate
case when one considers error-correcting codes\cite{s:ecc}. This
difference may be illustrated using figures~\ref{figure:phd3} and
\ref{figure:phdp}: the former is for $r<p$, and the ferromagnetic
phase is dominated by the 1RSB region in the limit $p \to \infty$ with
$r<p$; the latter is for $r=p$, and the ferromagnetic phase is
dominated by the RS region in the limit $r=p \to \infty$.

\begin{figure}
\begin{center}
\epsfig{file=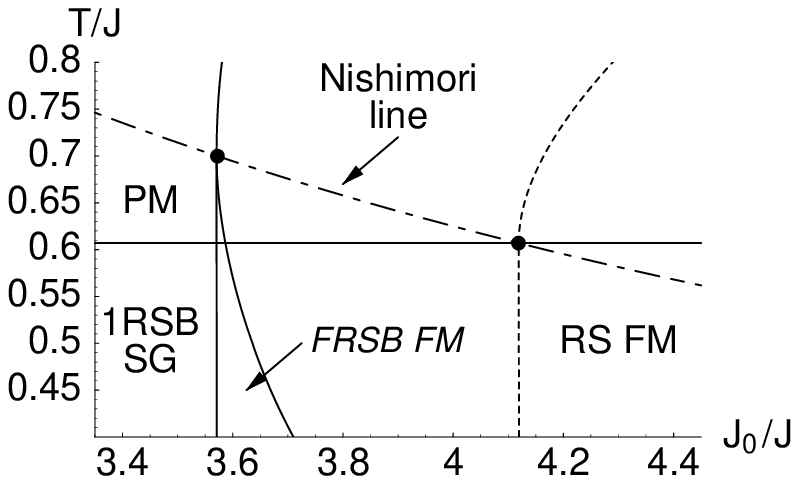,width=10cm}
\caption{A section of the phase diagram for the system given by the
Hamiltonian \eqref{Hfull} with $r=p=5$, as figure~\ref{figure:phdp},
showing the Nishimori line $p \beta J^2 = 2 J_0$ as a dot-dashed
line. The solution is replica symmetric with $M=q$ on this curve, as
discussed in \S\ref{sect:nish}. The lower-right point marks where the
paramagnetic, spin glass, and RS ferromagnetic phases meet in
thermodynamics. The upper-left point marks where the paramagnetic, the
RS ferromagnetic, and the FRSB ferromagnetic phases meet in
spinodals. The Nishimori line passes through both points.}
\label{figure:nish}
\end{center}
\end{figure}

\section{Special features on the Nishimori line\label{sect:nish}}

Some of our conclusions about the phase diagram for $r=p$ are
reinforced from a different point of view. We show in the present
section that on the Nishimori line, the distribution function of the
spin glass order parameter consists only of two delta functions. This
immediately implies that the ferromagnetic phase is not glassy on the
Nishimori line for any $r=p$.

We start our argument from the definition of the distribution function
of magnetization under the Hamiltonian \eqref{Hfull} with $r=p$:
\begin{gather}
P_m(y)=\left[ \frac{\sum_{\{ \sigma_i\}}
      \delta(y-N^{-1}\sum_i\sigma_i) \exp\left( -\beta \Ha_p \right)}
      {\sum_{\{ \sigma_i\}} \exp\left(-\beta \Ha_p\right)}
      \right]_J. \label{eq:m-def}
\end{gather}
Here the outer brackets denote the configurational average over the
Gaussian distribution of exchange interactions. To define the
distribution function of the spin glass order parameter, it is
convenient to introduce two replicas of the same system, with spins
$\{\sigma^{(1)}_i\}$ and $\{\sigma^{(2)}_i\}$:
\begin{gather}
    P_q(y)=\left[ \frac{
      \sum_{\{\sigma^{(1)}_i\}}\sum_{\{\sigma^{(2)}_i\}} \delta
      (y-N^{-1}\sum_i \sigma^{(1)}_i \sigma^{(2)}_i) \exp \left(-\beta
      \Ha_p(\{\sigma^{(1)}_i\})-\beta\Ha_p(\{\sigma^{(2)}_i\})\right)}
      {\sum_{\{\sigma^{(1)}_i\}}\sum_{\{\sigma^{(1)}_i\}}
      \exp\left(-\beta
      \Ha_p(\{\sigma^{(1)}_i\})-\beta\Ha_p(\{\sigma^{(2)}_i\})\right)}
      \right]_J. \label{eq:q-def}
\end{gather}
Our main result is the identity
\begin{gather}
     P_m(y)=P_q(y) \qquad \text{wherever} \qquad p \beta J^2 = 2 J_0.
     \label{eq:mq-identity}
\end{gather}
The condition $p\beta J^2=2 J_0$ defines a curve in the phase diagram
called the Nishimori line\cite{n:gauge}. This line passes through the
paramagnetic and ferromagnetic phases as shown by the dot-dashed curve
in figure~\ref{figure:nish}. The result is proved in \ref{app:gauge}.

It is well established that the distribution of magnetization $P_m(y)$
has a simple structure with two delta functions at symmetric
positions. Then \eqref{eq:mq-identity} proves that the distribution
function of the spin glass order parameter $P_q(y)$ is also simple on
the Nishimori line, which is not the case if the ferromagnetic phase
is glassy.

In \ref{app:morenish} we demonstrate three further facts about the
Nishimori line: it passes through the point where the spin glass
transition line meets the thermodynamic ferromagnetic transition line
(the lower-right point marked in figure~\ref{figure:nish}); at this
point, the $q_1$ of the spin glass and the $q$ of the ferromagnet are
equal; and it passes through the point where the continuous glassy
transition line (or, equivalently, the RS Almeida--Thouless line)
within the ferromagnetic phase meets the spinodal ferromagnetic
transition line (the upper-left point marked in
figure~\ref{figure:nish}). This last result demonstrates that there is
no D1RSB transition line within the ferromagnetic region at $r=p$, as
the continuous glassy transition reaches the spinodal line which marks
the edge of the ferromagnetic phases. It follows that $r_1 \leqslant
p$, as claimed in \S\ref{sect:res}.
\section{Conclusions}

In this paper, we have shown that the behaviour of a spin glass with
an $r$-spin ferromagnetic interaction of infinite range is simple to
obtain from a knowledge of the free energy either of the base spin
glass with a constrained magnetization, or equivalently of the base
spin glass in an external field (although these latter problems are
often numerically subtle); and we have applied this method to an Ising
glass with a $p$-spin Gaussian random interaction. We have found that,
for $p>2$ and $r \geqslant 2$, there are both glassy and non-glassy
ferromagnetic phases; and that for $r$ less than a $p$-dependent
critical value the glassy ferromagnet has regions of both one step and
full replica symmetry breaking, while for higher $r$ it has only full
replica symmetry breaking. We have also proved several exact results
for the case where $r=p$, which is relevant to error-correcting codes:
in particular, we have explicitly shown the existence of a curve of
simple form in the phase diagram, passing through both paramagnetic
and ferromagnetic phases, on which the solution is certainly
non-glassy. We hope that the present analysis will stimulate
investigation of the phase structures of related systems.

\appendix
\section{Results from perturbation theory close to the
Almeida--Thouless line\label{app:pert}}

In this appendix we derive expression for $x_\mathrm{c}$, the value of
$x$ along the C1RSB curve, and the condition to determine the RSB triple
point where the RS, 1RSB, and FRSB ferromagnetic phases meet. We use
the definitions
\begin{align}
\mathcal{T}_n &\doteq \int \frac{dz_0}{\sqrt{2\pi}}\, e^{-z_0^2/2}
\tanh^n \left( \sqrt{\tfrac{1}{2} p \beta^2 J^2 q_0^{p-1}} z_0 + \beta h
\right),
\\
\mathcal{S}_n &\doteq \int \frac{dz_0}{\sqrt{2\pi}}\, e^{-z_0^2/2}
\sech^n \left( \sqrt{\tfrac{1}{2} p \beta^2 J^2 q_0^{p-1}} z_0 + \beta h
\right).
\end{align}

A continuous one step replica symmetry breaking (C1RSB) transition
occurs on the line where $q_0=q_1$ in the 1RSB ansatz. Details of the
1RSB solution close to this line, using only the numerical solution to
the RS equation, are obtained by an expansion of these equations in
small $\epsilon \doteq (q_1-q_0)$. Since the hyperbolic functions of
$G$ become polynomials in $z_1$, the integrals over $z_1$ can be
carried out immediately, simplifying matters considerably. In
particular, one can find the RSB triple point using only the numerical
RS data.

To leading order, both \eqref{scrsb0} and \eqref{scrsb1} are
\begin{gather}
q_0 = \mathcal{T}_2 + O(\epsilon)
\label{pert1}
\end{gather}
and so $q_0 = q + O(\epsilon)$ where $q$ is the RS order parameter
satisfying \eqref{scrs}. To obtain a second piece of information from
these two equations, $[\eqref{scrsb1}-\eqref{scrsb0}]$ must be
expanded to first order, giving
\begin{gather}
\epsilon = \tfrac{1}{2} p(p-1) \beta^2 J^2 q_0^{p-2} \mathcal{S}_4\,
\epsilon + O(\epsilon)^2.
\label{pert2}
\end{gather}
This shows that \eqref{atrs} holds as an equality to leading order in
$\epsilon$, and hence that the C1RSB transition and the onset of RS AT
instability coincide. This is to be expected: where the replica
symmetry breaking mode of the RS solution softens, a continuous
transition to an RSB solution occurs.

\eqref{scrsbx} is trivially satisfied for $q_0=q_1$. Expanding to
first order gives \eqref{pert1} multiplied by $k_1 \doteq \tfrac{1}{4}
p(p-1) \beta^2 J^2 q_0^{p-2} \epsilon$, which gives no new
information. Expanding $[\eqref{scrsbx} - k_1 \eqref{scrsb0}]$ to
second order gives \eqref{pert1} multiplied by $k_2 \doteq
\tfrac{1}{8} p(p-1)(p-2) \beta^2 J^2 q_0^{p-3} \epsilon^2$ plus
\eqref{pert2} multiplied by $\tfrac{1}{2} k_1$, which again gives no
new information. Expanding $[\eqref{scrsbx} - (k_1+k_2) \eqref{scrsb0}
- \tfrac{1}{2} k_1 (\eqref{scrsb1}-\eqref{scrsb0})]$ to third order
gives
\begin{gather}
\tfrac{1}{12} \left[ \tfrac{1}{2} p (p-1) \beta^2 J^2 q_0^{p-2} \right]^3
\mathcal{S}_6 (x-x_\mathrm{c}) \epsilon^3 + O(\epsilon)^4 = 0
\label{pert3}
\\
\Rightarrow \qquad x = x_\mathrm{c} + O(\epsilon)
\intertext{where}
x_\mathrm{c} \doteq \frac{2 (p-2) q_0^3 + 4 p(p-1) \beta^2 J^2
q_0^{p+2} - 2 [p(p-1) \beta^2 J^2 q_0^p]^2 \mathcal{S}_6} {[p(p-1)
\beta^2 J^2 q_0^p]^2 \mathcal{S}_6}.
\label{pertxc}
\end{gather}
This gives a perturbative expression for $x$ on the C1RSB line using
only the RS value of $q$.\footnote{Our \eqref{pertxc} for
$x_\mathrm{c}$ differs from that given by (36) of de Oliveira and
Fontanari\cite{of:ipsgh}, due to their omission of one of these higher
order corrections. Further details may be found in our comment on that
paper\cite{gs:commof}.}

The calculation of the 1RSB stability of the C1RSB transition line
follows a similar method. Expanding \eqref{atrsb} in powers of
$\epsilon$, everything up to third order vanishes by the
self-consistent equations. Taking a linear combination of the
inequality \eqref{atrsb} and the equations \eqref{scrsb} such that all
terms to third order cancel, the fourth order term gives
\begin{multline}
4 (p-2) (8p-15) q + 48 p (p-1) (p-2) \beta^2 J^2 q^p \\ - 4 p^2
(p-1)^2 \beta^4 J^4 q^{2(p-1)} \left[ 4q + 3(p-2) (2+x_\mathrm{c})
\mathcal{S}_6 \right] \\ + 3 p^3 (p-1)^3 \beta^6 J^6 q^{3(p-1)} \left[
8 (1+x_\mathrm{c}) \mathcal{S}_6 - ( 6 + 8x_\mathrm{c} +
x_\mathrm{c}^2 ) \mathcal{S}_8 \right] > 0\,.
\label{pert4}
\end{multline}
At high temperatures, this is satisfied and the continuous replica
symmetry breaking transition is to 1RSB. Below a point (called the RSB
triple point) found by solving \eqref{scrs}, and \eqref{atrs} and
\eqref{pert4} as equalities, the transition is to FRSB. Within the RSB
phase there is a line marking a transition between 1RSB and FRSB which
meets the RS AT line at this triple point.

\section{An equation for the spinodal transition line \label{app:spin}}

In this appendix we use the definition
\begin{gather}
t \doteq \tanh \left( \sqrt{\tfrac{1}{2} p \beta^2 J^2 q^{p-1}} z +
\beta h \right).
\end{gather}

The spinodal transition to ferromagnetism occurs at the minimum $J_0$
for which an $M \neq 0$ solution is possible. In the RS region, the
ferromagnetic solution is given by \eqref{scrs}, \eqref{Mrs}, and
\eqref{heff}. Differentiating these three with respect to $h$, and
using integration by parts,
\begin{align}
\frac{\partial q}{\partial h} &= \int \frac{dz}{\sqrt{2\pi}}\,
e^{-z^2/2} \left[ (1-3t^2)(1-t^2) \tfrac{1}{2} p(p-1) \beta^2 J^2
q^{p-2} \frac{\partial q}{\partial h} + 2t (1-t^2) \beta \right],
\\
\frac{\partial M}{\partial h} &= \int \frac{dz}{\sqrt{2\pi}}\,
e^{-z^2/2} \left[ -t(1-t^2) \tfrac{1}{2} p(p-1) \beta^2 J^2 q^{p-2}
\frac{\partial q}{\partial h} + (1-t^2) \beta \right],
\\
1 &= \frac{\partial J_0}{\partial h} M^{r-1} + (r-1) J_0 M^{r-2}
\frac{\partial M}{\partial h}.
\end{align}
Asserting the spinodal condition $\partial J_0 / \partial h = 0$ and
eliminating the unwanted derivatives gives
\begin{gather}
M = (r-1) \beta h \left[ 1 - \int \frac{dz}{\sqrt{2\pi}}\, e^{-z^2/2}\,
t^2 - \frac{ 2 \left[ \int \frac{dz}{\sqrt{2\pi}}\, e^{-z^2/2}\, t
(1-t^2) \right]^2 } { \frac{2}{p(p-1) \beta^2 J^2 q^{p-2}} - \int
\frac{dz}{\sqrt{2\pi}}\, e^{-z^2/2}\, (1-t^2) (1-3t^2) } \right].
\label{spino}
\end{gather}

We are often interested in the intersection between the spinodal line
and the C1RSB line. At this point, \eqref{atrs} holds as an equality,
and \eqref{spino} simplifies to
\begin{gather}
M = (r-1) \beta h \left[ 1 - \int \frac{dz}{\sqrt{2\pi}}\,
e^{-z^2/2}\, t^2 - \frac{ \left[ \int \frac{dz}{\sqrt{2\pi}}\,
e^{-z^2/2}\, t (1-t^2) \right]^2 } { \frac{dz}{\sqrt{2\pi}}\,
e^{-z^2/2}\, t^2 (1-t^2) } \right].
\label{spinoc}
\end{gather}

This result can be used as discussed in \S\ref{sect:res}. The peak of
the 1RSB transition in the ferromagnetic region is given by the
solution of the RS self-consistency equation \eqref{scrs}, the RS AT
condition \eqref{atrs} as an equality, and the D1RSB equation
$x_\mathrm{c}=1$ with $x_\mathrm{c}$ given by \eqref{pertxc}. When
this peak hits the spinodal line, the D1RSB curve in the ferromagnetic
region vanishes. This occurs at $r=r_1$, obtained for given $p$ by
solving these three equations and \eqref{spinoc}. Similarly, the RSB
triple point is given by \eqref{scrs}, \eqref{atrs} as an equality,
and the perturbative 1RSB AT condition \eqref{pert4} as an
equality. When this point hits the spinodal line, the 1RSB
ferromagnetic region vanishes. This occurs at $r=r_2$, obtained by
solving these three equations and \eqref{spinoc}. Some values are
shown in table~\ref{tab:r}.

\section{Proofs for \S\ref{sect:nish}}

\subsection{Proof from gauge transformation\label{app:gauge}}

In this subsection, we prove \eqref{eq:mq-identity} using a gauge
transformation technique. We explicitly write the configurational
average appearing in \eqref{eq:m-def}:
\begin{align}
P_m(y) &= \int \prod_{i_1 <
\dots i_p} P_J(J_{i_1\dots i_p}) dJ_{i_1\dots i_p}\,
\frac{1}{Z_J(\beta)} \sum_{\{ \sigma_i\}} \delta \left(
y-N^{-1}\sum_i\sigma_i \right) \exp\left(-\beta
\Ha_p(\{\sigma_i\})\right)
\\
&= \int \mathcal{D}J\, \exp \left( -\frac{N^{p-1}}{p! J^2} \sum_{i_1 <
\dots i_p} J_{i_1\dots i_p}^2\right) \notag\\&\eqind\times
\frac{1}{Z_J(\beta)} \sum_{\{ \sigma_i\}} \delta \left(y-\tfrac{1}{N}
\sum_i\sigma_i\right) \exp \left[ -\beta \sum_{i_1 < \dots i_p} \left(
J_{i_1\dots i_p} - \frac{J_0 (p-1)!}{N^{p-1}} \right)
\sigma_{i_1}\dots \sigma_{i_p}\right]\notag\\&\eqind\times
\frac{1}{Z_J(\beta)} \sum_{\{ \tau_i \}} \exp \left[ -\beta \sum_{i_1
< \dots i_p} \left( J_{i_1\dots i_p} - \frac{J_0 (p-1)!}{N^{p-1}}
\right) \tau_{i_1}\dots \tau_{i_p}\right]
\intertext{where we have introduced a factor of unity at the end,
defined $\mathcal{D}J \doteq \prod_{i_1 < \dots i_p} 
\sqrt{\tfrac{N^{p-1}}{\pi p J^2}} dJ_{i_1 \dots i_p}$, and
$Z_J(\beta)$ is the partition function for a given choice of the
couplings $J_{i_1 \dots i_p}$:}
Z_J(\beta) &= \sum_{\{ \sigma_i \}} \exp \left[ -\beta \sum_{i_1 < \dots
i_p} \left( J_{i_1 \dots i_p} - \frac{J_0 (p-1)!}{N^{p-1}} \right)
\sigma_{i_1} \dots \sigma_{i_p}\right].
\intertext{Shifting the integration variable by $J_0(p-1)!/N^{p-1}$
and reordering trivially, we obtain}
P_m(y) &= \sum_{\{ \tau_i \}} \int \mathcal{D}J\, \exp \left(
-\frac{N^{p-1}}{p! J^2} \sum_{i_1 < \dots i_p} J_{i_1\dots i_p}^2 -
\sum_{i_1 < \dots i_p} \frac{2J_0}{pJ^2} J_{i_1 \dots i_p} - \sum_{i_1
< \dots i_p} \frac{J_0^2(p-1)!}{pN^{p-1}J^2} \right)
\notag\\&\eqind\times \frac{1}{Z_J(\beta)^2} \sum_{\{
\sigma_i\}} \delta \left(y-\tfrac{1}{N} \sum_i\sigma_i\right) \exp
\left[ -\beta \sum_{i_1 < \dots i_p} J_{i_1\dots i_p}
\sigma_{i_1}\dots \sigma_{i_p}\right] \notag\\&\eqind\times \exp
\left[ -\beta \sum_{i_1 < \dots i_p} J_{i_1\dots i_p} \tau_{i_1}\dots
\tau_{i_p}\right]\,.
\intertext{We now perform a gauge transformation with $\sigma_i \to
\sigma_i \tau_i$ and $J_{i_1 \dots i_p} \to J_{i_1 \dots i_p}
\tau_{i_1} \dots \tau_{i_p}$, and obtain}
P_m(y) &= \sum_{\{ \tau_i \}} \int \mathcal{D}J \, \exp \left(
-\frac{N^{p-1}}{p! J^2} \sum_{i_1 < \dots i_p} J_{i_1\dots i_p}^2 -
\sum_{i_1 < \dots i_p} \frac{2J_0}{pJ^2} J_{i_1 \dots i_p} \tau_{i_1}
\dots \tau_{i_p} - \sum_{i_1 < \dots i_p}
\frac{J_0^2(p-1)!}{pN^{p-1}J^2} \right) \notag\\&\eqind\times
\frac{1}{Z_J(\beta)^2} \sum_{\{ \sigma_i\}} \delta
\left(y-\tfrac{1}{N} \sum_i\sigma_i \tau_i \right) \exp \left[ -\beta
\sum_{i_1 < \dots i_p} J_{i_1\dots i_p} \sigma_{i_1}\dots
\sigma_{i_p}\right] \notag \\&\eqind\times \exp \left[ -\beta
\sum_{i_1 < \dots i_p} J_{i_1\dots i_p} \right]\,.
\intertext{If $2J_0/pJ^2 = \beta$ then exchanging two terms gives}
P_m(y) &= \sum_{\{ \tau_i \}} \int \mathcal{D}J \, \exp \left(
-\frac{N^{p-1}}{p! J^2} \sum_{i_1 < \dots i_p} J_{i_1\dots i_p}^2 -
\sum_{i_1 < \dots i_p} \frac{2J_0}{pJ^2} J_{i_1 \dots i_p} - \sum_{i_1
< \dots i_p} \frac{J_0^2(p-1)!}{pN^{p-1}J^2} \right)
\notag\\&\eqind\times \frac{1}{Z_J(\beta)^2} \sum_{\{ \sigma_i\}}
\delta \left(y-\tfrac{1}{N} \sum_i\sigma_i \tau_i \right) \exp \left[
-\beta \sum_{i_1 < \dots i_p} J_{i_1\dots i_p} \sigma_{i_1}\dots
\sigma_{i_p}\right] \notag \\&\eqind\times \exp \left[ -\beta
\sum_{i_1 < \dots i_p} J_{i_1\dots i_p} \tau_{i_1}\dots
\tau_{i_p}\right]\,.
\end{align}
Finally, shifting the integration variable back, we obtain
\begin{multline}
P_m(y) =
\\
\int \prod_{i_1 < \dots i_p} P_J(J_{i_1 \dots i_p}) dJ_{i_1
\dots i_p}\, \frac{1}{Z_J(\beta)^2} \sum_{\{
\sigma_i\}} \sum_{\{ \tau_i \}} \delta \left(y-\tfrac{1}{N}
\sum_i\sigma_i \tau_i \right) \exp \left[ -\beta \Ha_p(\{\sigma_i\}) -
\beta \Ha_p(\{\tau_i\}) \right]
\\
= P_q(y)
\end{multline}
which completes the proof.

\subsection{Other proofs concerning the Nishimori line\label{app:morenish}}

In this subsection, we prove the results concerning the Nishimori line
and two triple points mentioned in \S\ref{sect:nish}. (The
demonstration of several mathematical identities used here is left to
\ref{app:proofs}.) We first notice that \eqref{eq:mq-identity} means
not only that the ferromagnet on the Nishimori line ($p \beta J^2 = 2
J_0$) is replica symmetric, but that it has $q=M$. It follows that
$\tfrac{1}{2} p \beta^2 J^2 q^{p-1} = \beta J_0 M^{r-1}$, and with
\eqref{heff} this is $\beta h$. We may verify $q=M$ independently of
the gauge result, as using \eqref{scrs} and \eqref{Mrs} this becomes
\begin{gather}
\int \frac{dz}{\sqrt{2\pi}}\, e^{-z^2/2} \tanh^2 \left( \sqrt{\beta h}
z + \beta h \right) = \int \frac{dz}{\sqrt{2\pi}}\, e^{-z^2/2} \tanh
\left( \sqrt{\beta h} z + \beta h \right),
\label{pr1}
\end{gather}
an identity proved in \ref{app:proofs}.

The spin glass transition line satisfies \eqref{scrsb} with $x=1$ and
$q_0=h=0$ (as $M=0$). Matters are simplified by the independence of
$G$ from $z_0$, which means that the integrals over $z_0$ become
trivial. The assertion that the $q_1$ here is the same as the $q$ of
the RS ferromagnet on the Nishimori line becomes, with \eqref{scrsb1},
\begin{gather}
\int \frac{dz}{\sqrt{2\pi}}\, e^{-z^2/2} \tanh^2 \left( \sqrt{\beta h}
z + \beta h \right) = \frac{ \int \frac{dz_1}{\sqrt{2\pi}}\,
e^{-z_1^2/2} \cosh \left( \sqrt{\beta h} z_1 \right) \tanh^2 \left(
\sqrt{\beta h} z_1 \right) } { \int \frac{dz_1}{\sqrt{2\pi}}\,
e^{-z_1^2/2} \cosh \left( \sqrt{\beta h} z_1 \right) },
\label{pr2}
\end{gather}
which is also proved in \ref{app:proofs}. The other non-trivial
self-consistent equation, \eqref{scrsbx}, becomes
\begin{multline}
\frac{1}{4} (p-1) \beta^2 J^2 q^p = - \ln \int
\frac{dz_1}{\sqrt{2\pi}}\, e^{-z_1^2/2} \cosh \left( \sqrt{\beta h}
z_1 \right) \\ + \frac{ \int \frac{dz_1}{\sqrt{2\pi}}\, e^{-z_1^2/2}
\cosh \left( \sqrt{\beta h} z_1 \right) \ln \cosh \left( \sqrt{\beta
h} z_1 \right) } { \int \frac{dz_1}{\sqrt{2\pi}}\, e^{-z_1^2/2} \cosh
\left( \sqrt{\beta h} z_1 \right)}.
\label{scsgx}
\end{multline}

The thermodynamic transition line is located by equating the free
energy of the ferromagnetic solution and that of the paramagnetic or
spin glass solution. The free energy for $r=p$ is related to that in a
field via \eqref{genfr} and \eqref{genf1}. Where the thermodynamic and
spin glass transition lines cross, the free energy of the spin glass
is $f_\mathrm{SG} = \ffRSB$ with $x=1$, $q_0=h=0$, and $\ffRSB$ given
by \eqref{fRSB}. If the Nishimori line passes through this point then
the previous results apply. The free energy of the ferromagnet is
therefore
\begin{gather}
f_\mathrm{FM} = \ffRS + h M - \tfrac{1}{r} J_0 M^r = \ffRS +
\tfrac{1}{2} (p-1) \beta J^2 q^p
\end{gather}
with $\ffRS$ given by \eqref{fRS}. The thermodynamic transition
condition $f_\mathrm{FM}=f_\mathrm{SG}$ becomes
\begin{gather}
\frac{1}{4} (p-1) \beta^2 J^2 q^p - \int \frac{dz}{\sqrt{2\pi}}\,
e^{-z^2/2} \ln \cosh \left( \sqrt{\beta h} z + \beta h \right) = - \ln
\int \frac{dz_1}{\sqrt{2\pi}}\, e^{-z_1^2/2} \cosh \left( \sqrt{\beta
h} z_1 \right)
\intertext{or, using \eqref{scsgx},}
\frac{ \int \frac{dz_1}{\sqrt{2\pi}}\, e^{-z_1^2/2}
\cosh \left( \sqrt{\beta h} z_1 \right) \ln \cosh \left( \sqrt{\beta
h} z_1 \right) } { \int \frac{dz_1}{\sqrt{2\pi}}\, e^{-z_1^2/2} \cosh
\left( \sqrt{\beta h} z_1 \right)} - \int \frac{dz}{\sqrt{2\pi}}\,
e^{-z^2/2} \ln \cosh \left( \sqrt{\beta h} z + \beta h \right) = 0\,,
\label{pr3}
\end{gather}
which is also proved in \ref{app:proofs}. Thus, the Nishimori line
does pass through the thermodynamic triple point where the paramagnet,
spin glass, and RS ferromagnets meet.

The equations for the intersection of spinodal line and the C1RSB
transition line are discussed in \ref{app:spin}. If this point lies on
the Nishimori line then the above results apply. Setting $p=r$, $M=q$,
and $\tfrac{1}{2} p \beta^2 J^2 q^{p-1} = \beta h$ in \eqref{spinoc}
and in \eqref{atrs} as an equality, dividing the former by the latter,
and dividing the result by $\tfrac{1}{2} p(p-1) \beta^2 J^2 q^{p-1}$,
gives a condition for the consistency of this hypothesis which may be
written
\begin{multline}
\int \frac{dz}{\sqrt{2\pi}}\, e^{-z^2/2} \sech^4 \left( \sqrt{\beta h}
z + \beta h \right) - 1 + \int \frac{dz}{\sqrt{2\pi}}\, e^{-z^2/2}
\tanh^2 \left( \sqrt{\beta h} z + \beta h \right) \\ + \frac{ \left[
\int \frac{dz}{\sqrt{2\pi}}\, e^{-z^2/2} \tanh \left( \sqrt{\beta h} z
+ \beta h \right) \sech^2 \left( \sqrt{\beta h} z + \beta h \right)
\right]^2 } { \int \frac{dz}{\sqrt{2\pi}}\, e^{-z^2/2} \tanh^2 \left(
\sqrt{\beta h} z + \beta h \right) \sech^2 \left( \sqrt{\beta h} z +
\beta h \right) } = 0\,,
\label{pr4}
\end{multline}
which is proved in \ref{app:proofs}. Thus, the Nishimori line also
passes through the spinodal triple point where the paramagnet, RS
ferromagnet, and RSB ferromagnet meet.

\subsection{Proofs of some mathematical identities\label{app:proofs}}

In this section we prove several identities used in
\ref{app:morenish}. These proofs will all rely on the lemmas
\begin{align}
\int \frac{dz}{\sqrt{2\pi}}\, e^{-z^2/2} \cosh(az) f(z) &= e^{a^2/2}
\int \frac{dz}{\sqrt{2\pi}}\, e^{-z^2/2} \fev(z+a)
\label{lemev}
\\
\int \frac{dz}{\sqrt{2\pi}}\, e^{-z^2/2} \sinh(az) f(z) &= e^{a^2/2}
\int \frac{dz}{\sqrt{2\pi}}\, e^{-z^2/2} \fod(z+a)
\label{lemod}
\end{align}
where $\fev$ and $\fod$ are the even and odd parts of the function
$f$. The proof of these runs as follows:
\begin{align}
\int \frac{dz}{\sqrt{2\pi}}\, e^{-z^2/2} \left\{ \cosh(az) \atop
\sinh(az)\right\} f(z) &= \frac{1}{2} \int \frac{dz}{\sqrt{2\pi}}\,
e^{-z^2/2} \left( e^{az} \pm e^{-az} \right) f(z)
\\
&= \frac{1}{2} e^{a^2/2} \int \frac{dz}{\sqrt{2\pi}}\, \left(
e^{-(z-a)^2/2} \pm e^{-(z+a)^2/2} \right) f(z)
\\
&= \frac{1}{2} e^{a^2/2} \left( \int \frac{dz_1}{\sqrt{2\pi}}\,
e^{-z_1^2/2} f(z_1+a) \pm \int \frac{dz_2}{\sqrt{2\pi}}\, e^{-z_2^2/2}
f(-z_2-a) \right)
\\
&= e^{a^2/2} \int \frac{dz}{\sqrt{2\pi}}\, e^{-z^2/2} \left\{ \fev(z+a)
\atop \fod(z+a) \right\}.
\end{align}

The proof of \eqref{pr1} now runs as follows:
\begin{align}
\int \frac{dz}{\sqrt{2\pi}}\, e^{-z^2/2} \tanh^2 (az+a^2) &=
e^{-a^2/2} \int \frac{dz}{\sqrt{2\pi}}\, e^{-z^2/2} \cosh(az)
\tanh^2(az) &\text{using \eqref{lemev}}
\\
&= e^{-a^2/2} \int \frac{dz}{\sqrt{2\pi}}\, e^{-z^2/2} \sinh(az)
\tanh(az)
\\
&= \int \frac{dz}{\sqrt{2\pi}}\, e^{-z^2/2} \tanh(az+a^2) &
\text{using \eqref{lemod}.}
\end{align}

The proof of \eqref{pr2} runs as follows:
\begin{align}
\int \frac{dz}{\sqrt{2\pi}}\, e^{-z^2/2} \cosh(az) &= e^{a^2/2} \int
\frac{dz}{\sqrt{2\pi}}\, e^{-z^2/2} &\text{using \eqref{lemev}}
\\
&= e^{a^2/2}\,;
\\
\int \frac{dz}{\sqrt{2\pi}}\, e^{-z^2/2} \cosh(az) \tanh^2(az) &=
e^{a^2/2} \int \frac{dz}{\sqrt{2\pi}}\, e^{-z^2/2} \tanh^2(az+a^2)\,;
&\text{using \eqref{lemev}}
\\
\frac{ \int \frac{dz}{\sqrt{2\pi}}\, e^{-z^2/2} \cosh(az) \tanh^2(az)
} { \int \frac{dz}{\sqrt{2\pi}}\, e^{-z^2/2} \cosh(az) } &= \int
\frac{dz}{\sqrt{2\pi}}\, e^{-z^2/2} \tanh^2(az+a^2)\,,
\end{align}
and the proof of \eqref{pr3} is identical with $\tanh^2$ replaced by
$\ln\cosh$.

The proof of \eqref{pr4} runs as follows: using \eqref{lemev},
\begin{align}
\int \frac{dz}{\sqrt{2\pi}}\, e^{-z^2/2} \sech^4 (az+a^2) &=
e^{-a^2/2} \int \frac{dz}{\sqrt{2\pi}}\, e^{-z^2/2} \cosh(az)
\sech^4(az)\,,
\\
1 &= e^{-a^2/2} \int \frac{dz}{\sqrt{2\pi}}\, e^{-z^2/2} \cosh(az)\,,
\\
\int \frac{dz}{\sqrt{2\pi}}\, e^{-z^2/2} \tanh^2 (az+a^2) &=
e^{-a^2/2} \int \frac{dz}{\sqrt{2\pi}}\, e^{-z^2/2} \cosh(az)
\tanh^2(az)\,;
\intertext{using \eqref{lemod},}
\int \frac{dz}{\sqrt{2\pi}}\, e^{-z^2/2} & \tanh(az+a^2) \sech^2(az+a^2)
\notag\\ &= e^{-a^2/2} \int \frac{dz}{\sqrt{2\pi}}\, e^{-z^2/2} \sinh(az)
\tanh(az) \sech^2(az)
\\
&= e^{-a^2/2} \int \frac{dz}{\sqrt{2\pi}}\, e^{-z^2/2} \cosh(az)
\tanh^2(az) \sech^2(az)\,;
\intertext{and using \eqref{lemev} again}
\int \frac{dz}{\sqrt{2\pi}}\, e^{-z^2/2} & \tanh^2(az+a^2)
\sech^2(az+a^2) \notag\\&= e^{-a^2/2} \int \frac{dz}{\sqrt{2\pi}}\,
e^{-z^2/2} \cosh(az) \tanh^2(az) \sech^2(az)\,;
\end{align}
thus
\begin{multline}
\int \frac{dz}{\sqrt{2\pi}}\, e^{-z^2/2} \sech^4(az+a^2) - 1 + \int
\frac{dz}{\sqrt{2\pi}}\, e^{-z^2/2} \tanh^2(az+a^2) \\ + \frac{ \left[
\int \frac{dz}{\sqrt{2\pi}}\, e^{-z^2/2} \tanh(az+a^2) \sech^2(az+a^2)
\right]^2 } { \int \frac{dz}{\sqrt{2\pi}}\, e^{-z^2/2} \tanh^2(az+a^2)
\sech^2(az+a^2) } \\ = e^{-a^2/2} \int \frac{dz}{\sqrt{2\pi}}\,
e^{-z^2/2} \cosh(az) \left[ \sech^4(az) - 1 + \tanh^2(az) +
\tanh^2(az) \sech^2(az) \right];
\end{multline}
and this is identically zero since the term in square brackets
vanishes using $\sech^2(az) = 1 - \tanh^2(az)$.

\section*{Acknowledgments}

PG and DS would like to thank EPSRC (UK) for financial support, PG for
research studentship 97304251 and DS for research grant GR/M04426. HN
acknowledges the support of the Sumitomo Foundation. This work was
also supported by the Anglo--Japanese Collaboration Programme between
The Royal Society and the Japan Society for the Promotion of Science.

\end{document}